\documentclass[prl, twocolumn, showpacs, amsmath]{revtex4-1}
\usepackage{dcolumn}
\usepackage{bm}
\usepackage{graphicx}
\usepackage{color}
\usepackage{epsfig}
\usepackage{amsmath}
\usepackage{amssymb}

\DeclareGraphicsExtensions{. jpg,. pdf, . mps, . png, . eps, . ps, . EPS}

\begin{document}
\def\be{\begin{equation}}
\def\ee{\end{equation}}
\newcommand{\grphi}[1]{{\boldsymbol \nabla} \phi}

\def\bc{\begin{center}} 
\def\ec{\end{center}}
\def\bea{\begin{eqnarray}}
\def\eea{\end{eqnarray}}
\newcommand{\avg}[1]{\langle{#1}\rangle}
\newcommand{\Avg}[1]{\left\langle{#1}\right\rangle}

\title{The Generalized Arrhenius law in out of equilibrium systems}

\author{Serena Bradde$^1$ and Giulio Biroli$^2$}
\affiliation{$^1$ Computational Biology program, Memorial Sloan-Kettering Cancer Center, New York, U.S.A and \\Physics of Biological Systems, Institut Pasteur, 75724, Paris Cedex 15, France\\
$^2$Institut de Physique Th\'eorique (IPhT) CEA Saclay and CNRS URA 2306, 91191 Gif Sur Yvette, France}

\begin{abstract}
In this work we provide a comprehensive analysis of the activation problem out of equilibrium. 
We generalize the Arrhenius law for systems driven by non conservative time independent forces, subjected to retarded friction and non-Markovian noise. The role of the energy function is now played by the out of equilibrium potential $\phi=-\lim_{T\rightarrow 0}T\log P_s$, with $P_s$ being the steady state probability distribution and $T$ the strength of the noise. We unveil the relationship
between the generalized Arrhenius law and a time-reversal transformation discussed in the context of fluctuations theorems out of equilibrium. Moreover, we characterize the noise-activated trajectories by obtaining their explicit expressions and identifying their irreversible nature. Finally, we discuss a real biological application that illustrates our results. 
\end{abstract}
\pacs{05. 40. -a, 05. 70. -a, 05. 10. Gg} 

\maketitle
The Arrhenius law originally discovered for the rate of chemical reactions \cite{Achemistry}
is one of the most important principles governing the behavior of systems 
characterized by several energy scales.
The diffusivity of vacancies in 
crystals \cite{Avacancy}, the viscosity of strong super-cooled liquids \cite{bbreview}, 
the rate of protein folding \cite{Aprotein} 
are just few examples showing the broad range of physical situations in which the Arrhenius law holds.
This happens whenever the dynamics of the system at hand is dominated 
by a process requiring an activation energy $\Delta E$ much larger than the temperature. 
In this case, the time-scale characterizing the dynamical behavior is given by 
\[
\tau\simeq \tau_0 \exp\left(\frac{\Delta E}{T} \right)\,,
\]
thus leading to the same behavior for several observables related to the relaxation time (rates, diffusion coefficients, etc.). In this equation $\tau_0$ is, roughly speaking, the high-temperature time-scale \footnote{For simplicity of notations we henceforth set $k_B=1$.}.

The usual domain of applicability of the Arrhenius law is restricted to systems at 
{\it equilibrium} (or 
in a metastable equilibrium, as in the case of nucleation processes). In this case
the Arrhenius law can be derived in several ways. The original proof is due to Kramers \cite{RevModPhys}, who
obtained it for a particle undergoing Langevin dynamics with white noise. 
A natural question is to what extent this can be generalized to non-equilibrium systems, which are nowadays witnessing a 
growing interest, e.g. 
gently shaken granular media, sheared fluids at low temperature, active matter, single molecule 
experiments, biochemical networks. These systems are out of equilibrium because are driven by forces that do not
derive from a potential and are subjected to noise that does not necessarily correspond to a thermal bath at equilibrium. 
If the force field is time-independent then a stationary probability 
distribution is generally reached at long time, but it is not given by the Gibbs-Boltzmann distribution $\exp(-E/T)$ since for non-conservative force fields there is no energy function to start with. In consequence, the usual intuition behind the Arrhenius law, 
which is based on a stochastic dynamics in an energy landscape characterized by a barrier $\Delta E$, becomes meaningless.
Yet, experiments, such the ones of \cite{annagremaud} on shaken granular media, suggest that in the low noise regime a generalization of this law does exist.
Despite some results obtained in specific contexts \cite{kanebray,borgis,Gardiner,joanny,chertkov,RevModPhys,Wang2008}, a general and comprehensive analysis addressing this issue is still lacking. Here we fill this gap: for driven systems subjected to non-thermal and in general multiplicative non-Markovian noise we derive the generalization of the 
Arrhenius law and identify the corresponding rare noise-activated paths followed during the dynamics. Our theory applies to force fields characterized by more than one stable attractor and in the small noise limit, meaning that the timescale to "jump" from one attractor to the other is much larger than 
the characteristic timescales to vibrate around them. As we shall explain, our results 
can be understood physically in terms of a generalization of time-reversal symmetry discovered in the context of fluctuations theorems \cite{hatanosasa,bertini,jorgelectures}. 
In order to illustrate our results we provide an explicit example inspired by bacteria evolution in the intestine caused by antibiotic administration \cite{ploscompbio}. 
 
We start our analysis by focusing on a simplified model defined by the Langevin equation:
\begin{equation}
\gamma \frac{d{\mathbf x}}{dt}={\bf F}+{\boldsymbol \eta}
\end{equation}
where ${\bf F}$ is the force field, $\gamma$ is the viscosity, ${\boldsymbol \eta}$ a Gaussian multiplicative noise, possibly corresponding to thermal fluctuations. We assume that ${\boldsymbol \eta}$ has zero mean and variance $\langle\eta_\alpha(t)\eta_\beta(t')\rangle=2T \gamma H({\mathbf x}) \delta(t-t')\delta_{\alpha,\beta}$ where $H({\mathbf x})$ is a generic positive function of ${\mathbf x}$ and
$T$ is the strength of the noise fluctuations (it only coincides with temperature for an equilibrated bath). The force field ${\bf F}$ is non-conservative therefore the system is kept out of equilibrium: 
it dissipates heat with the reservoir and does (or receives) work because of the force field. 
More general systems containing an inertial term, retarded friction and non-Markovian noise will be considered afterward.  
In order to study a simple but instructive case of activated non-equilibrium dynamics we focus on a force field characterized by three fixed points 
corresponding to ${\bf F}=0$ (two stable, $S_1, S_2$, and one unstable $U$). 
This is the counterpart of the usual process, a jump through an energy barrier, studied to derive
the "equilibrium" Arrhenius law. 
Our derivation is based on the Martin-Siggia-Rose field theory \cite{MSR}, whose action using Ito convention is: 
\[S=-\int_{t_i}^{t_f} dt' \hat {\bf x} \cdot \left(\frac{d{\bf x}}{dt}-{\bf F}({\bf x}) -T H({\bf x}) \hat{\bf x}\right)\]  
where $\hat {\bf x}$ is the response field and we use time units such that $\gamma=1$.\\ 
In the low noise limit, $T\rightarrow 0$, the dynamics can be described in the following way. The system first flows following the force field toward one of the stable points, whether this is 
$S_1$ or $S_2$ depends on the starting point, and then fluctuates locally around it. On much larger times, rare configurations of the noise eventually induce far away excursions allowing the system to escape from one basin of attraction to the other.  
The characterization of these dynamical processes can be obtained by computing the probability $P(S_1\rightarrow S_2;t)$ that a system starting in $S_1$ at time $t_i$ is in $S_2$ at a very large time $t_f=t+t_i$. This is given by  
sum over all paths $[{\bf x}(t)]$ that connect these two points in a time $t$. Each trajectory is
weighted by the exponential of its corresponding action. 
In the low noise
limit it is straightforward to check that the sum over paths is dominated by the saddle point contributions because one can pull out a $1/T$ factor in front of $S$ by 
redefining $\hat {\bf x}\rightarrow \hat {\bf x}/T$. The functional integral is thus approximated by the sum of
the paths that verify:
\begin{eqnarray}
\label{MNsolDxhat}&&\frac{\delta S}{\delta x_\alpha}=0 :\quad\frac{d\hat{x}_\alpha}{dt}=-\sum_\beta \left[\frac{\partial F_\beta}{\partial x_\alpha}\hat{x}_\beta + T\hat{x}^2_\beta \frac{\partial H({\bf x})}{\partial x_\alpha}\right]\nonumber\\
\label{MNsolDx}&&\frac{\delta S}{\delta \hat{x}_\alpha}=0 :\quad\frac{dx_\alpha}{dt}-F_\alpha=2TH({\bf x})\hat{x}_\alpha
\end{eqnarray}
In the equilibrium case, the solution of the saddle point equations can be constructed in terms of the downhill trajectory, 
that corresponds to the path going from the unstable point towards one of the stable ones, and the uphill trajectory
that corresponds to a path going in the opposite direction. The extremal paths on which one has to sum 
to obtain  $P(S_1\rightarrow S_2;t)$ are all possible combinations of the uphill and the downhill ones concatenated in such a way to verify the boundary conditions ${\bf x}(t_i)={\bf x}_{S_1}$ and ${\bf x}(t_f)={\bf x}_{S_2}$. The detailed computation, which 
is analogous to the so called dilute instanton gas, is presented in \cite{CCR}. In order to extend
it to the non-equilibrium case we look for the generalization of the downhill and the uphill solutions. 
The former is immediately found: it corresponds to $\dot {\bf x}={\bf F}$ and $\hat{\bf x}=0$ and leads to a null action. This means that the corresponding weight is of the order of one, as expected  
physically, since the downhill trajectory is a typical one and does not need to be activated by the noise. 
The main problem is to find the uphill trajectory going, say, from $S_1$ to $U$. 
In equilibrium conditions, corresponding to ${\bf F}=-{\boldsymbol \nabla} E$ and $H=1$, the uphill trajectory is the time-reversed of the downhill one and reads $\dot{\bf x}={\boldsymbol \nabla} E$ and $\hat{\bf x}= \dot{\bf x}/T={\boldsymbol \nabla} E/T$. 
The corresponding action $S=-(E(U)-E(S_1))/T=-\Delta E/T$ leads to an \emph{Arrhenius} weight, $e^{-\Delta E/T}$, for the uphill solution.\\
The naive guess for the generalization to the non-equilibrium case where one substitutes $ -\boldsymbol{\nabla} E$ with  $\bf{F} $, leading to $\dot{\bf x}=-{\bf F}$ and $\hat{\bf x}=\dot{\bf x}/T$, does not work. 
In order to find the solution of this conundrum, we have to introduce the zero noise non-equilibrium potential $\phi=\lim_{T\rightarrow 0}-T\log P_s({\bf x})$, with $P_s$ being the stationary probability distribution. In mathematical language  $\phi$ is the large deviation function 
determining the probability of rare events in the zero noise limit.  Plugging $P_s\simeq e^{-\phi/T}$ into the Fokker-Planck 
equation one finds that $\phi$ verifies the equation
\begin{equation}\label{eqphi}
{\boldsymbol \nabla} \phi \cdot ( {\boldsymbol F}+H({\bf x}) {\boldsymbol \nabla} \phi ) +  \mathcal{O}\left(T\right)=0
\end{equation}
Since $\phi$ somehow plays the role of the energy function, 
a natural generalization of the uphill solution can be found replacing $E$ with $\phi$ in the expression for 
$\hat{\bf x}$, {\it i.e.} $\hat{\bf x}={\boldsymbol \nabla} \phi/T$. Given the second equation of (\ref{MNsolDx}), this choice leads to:
\begin{equation}\label{uphill}
\dot {\bf x}={\bf F}+2H({\bf x})\grphi{}\qquad \hat{\bf x}=\grphi{}  /T \,,
\end{equation}
which indeed provides a solution also for first equation of (\ref{MNsolDxhat}), as it can be checked
by direct substitution and by using the derivative of eq.~(\ref{eqphi}). 
The action associated to the uphill solution reads:
\begin{equation}
S=-\frac{1}{T}\int dt' \grphi{}\cdot  \dot {\bf x}+ \grphi{}\cdot ({\bf F}+H({\bf x})\grphi{}) 
\end{equation}
Noticing that the last term between parenthesis is zero because of (\ref{eqphi}), we are left with the integral of a total derivative and, therefore, 
$S=-\frac{\Delta \phi}{T}$. Hence, we indeed obtain the generalized \emph{Arrhenius} weight, $e^{-\Delta \phi/T}$, for the uphill solution. 
By combining together uphill and downhill solutions, as done in \cite {CCR}, one finds that 
\begin{eqnarray}
&&P(S_1\rightarrow S_2;t)=P_{s}({\bf x}_{S_2})\left(1-e^{-\frac{t}{\tau}} \right) 
\nonumber\\
&&\tau(T)\simeq\tau_0\exp\left(\frac{\Delta \phi}{T}\right)\nonumber
\end{eqnarray}
where $\tau_0$ contains the sub-leading contributions to the generalized Arrhenius law 
coming from the Gaussian fluctuations around the instantons. On the basis of the results of 
\cite{borgis} we expect that an explicit computation should lead to the result
$\tau_0^{-1}=|\lambda_0|/(2\pi) |\prod_i(\omega_i^U)/(\omega_i^{S_1})|^{1/2}$
where $\lambda_0$ is the real part of the lowest eigenvalue of $\partial F_{\alpha}/\partial x_{\beta}$ evaluated in $U$
and the $\omega_i^U$s and the $\omega_i^{S_1}$s are the eigenvalues of the Hessian of $\phi$ evaluated in
$U$ and $S_1$ respectively. \\
Note that in general the solution of a system of 
equations like (\ref{MNsolDx}) with fixed initial and final conditions is not straightforward at all, and 
can be only obtained numerically by the shooting method (one searches by trial and error the initial position and velocity that leads to the correct solution). Thus, being able to provide the explicit solution and, in top of that, 
finding that the action is the integral of a total derivative are quite unexpected simplifications. In equilibrium, they are due to the existence of the time-reversal symmetry \cite{jorgelectures}. Remarkably, a generalization of the time-reversal transformation, which was introduced
in the context of fluctuations theorems out of equilibrium \cite{hatanosasa,bertini,jorgelectures,chertkov}, 
provides a physical explanation for the non-equilibrium case too. 
This transformation gives a relationship between the probability of a dynamical path and its time reversal in the so called adjoint dynamics corresponding to the Langevin evolution in a {\em renormalized} force field 
${\bf F}_A=-{\bf F}-2H({\bf x})\grphi{}$: 
\begin{equation}\label{TRS}
P([{\bf x}(t)];S_1\rightarrow U)=P_A([{\bf x}(-t)];U\rightarrow S_1)\;e^{-\frac{\Delta\phi}{T} }\,.
\end{equation}
In this expression $P_A$ denotes the probability of paths evolving with the adjoint dynamics \footnote{This relationship is more general and valid also at finite temperature if one replaces $\phi$ with $-T \log P_s$. Moreover, the initial and final point do not need to be stationary points.}. 
In order to identify the uphill trajectory one has to maximize the LHS. Instead of solving this difficult problem, one can solve the much easier one consisting in maximizing the RHS associated to the downhill trajectory in the adjoint dynamics. Since the exponential term is path-independent (the boundary conditions are fixed), the most probable path for $T\rightarrow 0$ is just the one that follows the force field ${\bf F}_A=-{\bf F}-2H({\bf x})\grphi{}$ going from the unstable point $U$ to the stable one $S_1$
\footnote{Stable (unstable) points with the standard dynamics remain stable (unstable) points for the adjoint dynamics. This can be proved by showing that the Hessian matrices at the critical points for the standard dynamics and its adjoint have the same eigenvalues.}. Because of the identity (\ref{TRS}), the time-reversal of this path is the one maximizing the LHS and, indeed, corresponds to the uphill solution.
In conclusion, the generalization of the time reversal transformation provides the physical principle behind the generalization of the Arrhenius law and an explanation for the unexpected simplifications we discovered. 
 
An important difference with the equilibrium case is that the large deviation function $\phi$ is not known {\it a priori}. Thus, it may seem that the previous results are of little use in practice. However, the explicit integration of eqs. in (\ref{MNsolDx}) between 
$S_1$ (or $S_2$) and $U$ and the evaluation of the corresponding action allows one to explicitly obtain $\Delta \phi$. Actually, this procedure is more general and allows one to completely reconstruct $\phi({\bf x})$ up to a normalization constant.  In order to do it, one can partition the phase space in basins of attraction associated to each stable point. The non-equilibrium potential for a point belonging to the basin of, say, $S_1$ is given by $\phi({\bf x}_{S_1})$ plus the action evaluated on the trajectory that connects $S_1$ to this point. This can be understood by noticing that under the time-reversal mapping the action $S$ transforms into the action for the downhill solution, which is null, plus $\Delta \phi/T$ between the final and the initial points. By comparing the values of $\Delta \phi$ for points along the separatrix one can completely reconstruct $\phi({\bf x})$ up to a normalization constant. Mathematically, this is related to an underlying Hamilton-Jacobi theory that applies to eqs. like the ones in (\ref{MNsolDx}) \cite{bertini,graham}. 

In order to show an explicit example of our theoretical framework, we focus on a biological-inspired example related to 
the study of the time evolution induced by an antibiotic treatment of bacterial microbes living in the human intestine \cite{ploscompbio}.
This dynamical system was modeled in terms of an  over-damped Langevin equation characterized by white noise and the non-conservative force field
$\mathbf{F}(x,y)=(\frac{x}{x+fy}-\epsilon x,\frac{f y}{x+fy} -\psi x y -y)$. 
The variables $x$ and $y$ describe the concentrations of the bacteria which are respectively sensitive and 
tolerant to the antibiotic. The constant $\epsilon, f$ and $\psi $ are related to mortality rate, fitness and interactions between bacteria
populations. This system exhibits a region of bistability for a particular choice of the parameteres with two stable points $S_1=(0,1)$ and $S_2=(1/\epsilon,0)$, and an unstable one $U=(\frac{\epsilon f-1}{\psi},\frac{\psi+\epsilon(1-\epsilon f)}{\psi f\epsilon})$. 
In order to illustrate our theory, we compute the non-equilibrium potential $\phi$ by evaluating the action 
for all the extremal trajectories starting from stable points (Fig.1).
It is double-well shaped with two minima associated to $S_1$ and $S_2$ and one maximum in correspondence
to $U$.
\begin{figure}
\includegraphics[width=.5\textwidth]{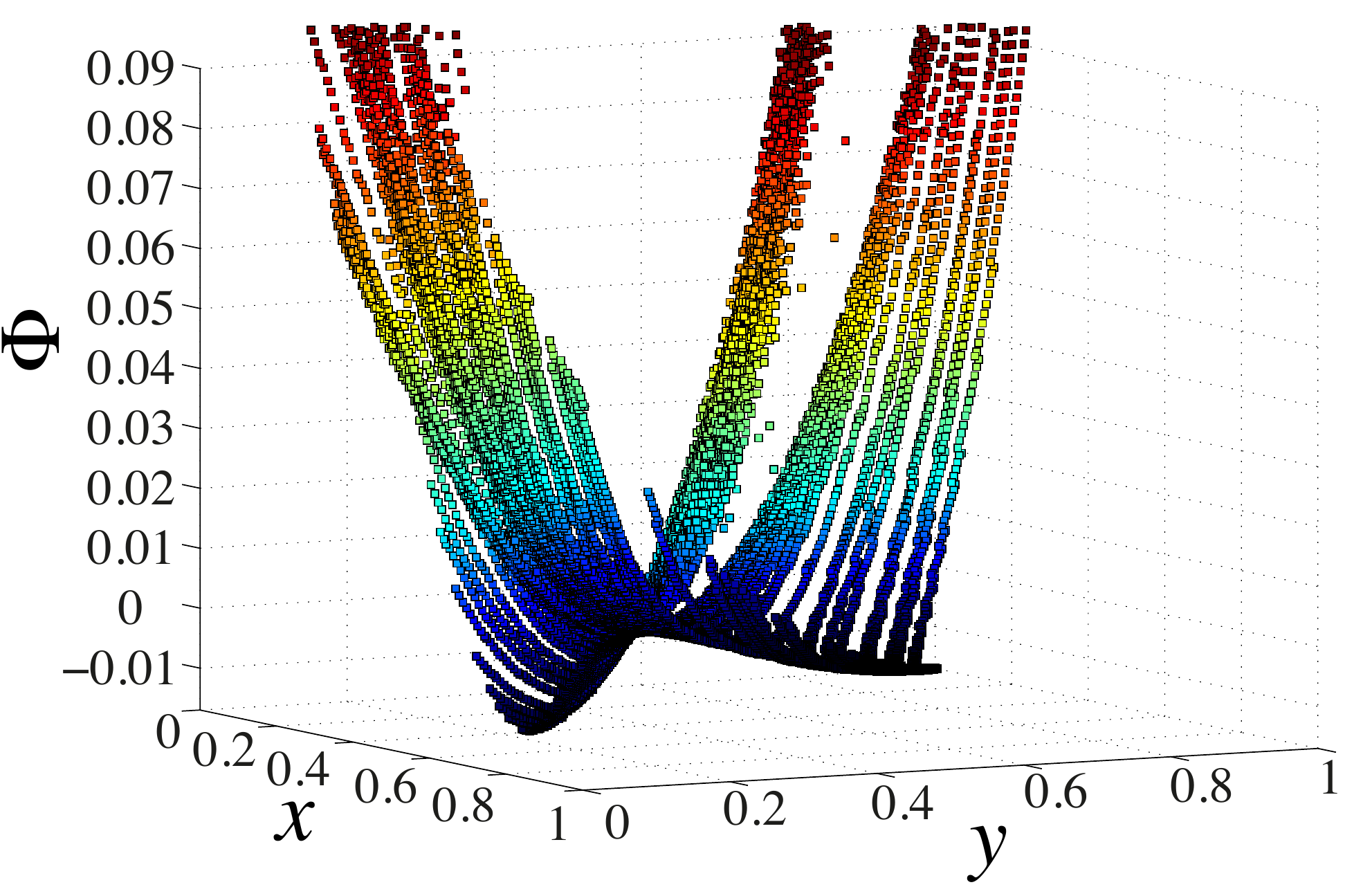}
\caption{Non-equilibrium potential for $\epsilon=1.2$, $f=1.12$,  $\psi=0.7$.}
\label{potential}
\end{figure}
Then we characterize the optimal paths that interpolates between the two stable points, see Fig.\ref{saddlepoints}. The 
most striking difference with the equilibrium case is that uphill and downhill trajectories pass through different regions of the phase space. 
The reason for this can be understood by recalling that out-of-equilibrium systems substain a steady state probability current, ${\bf J}=({\bf F}-T{\boldsymbol \nabla})P_s$. In the zero noise limit ${\bf J}$ equals $({\bf F}+\grphi{})P_s$ and is orthogonal everywhere to the gradient of the potential because of eq. (\ref{eqphi}). Hence, uphill and downhill trajectories can be respectively decomposed in two perpendicular contributions: $\left.\dot {\bf x}\right|_{uphill}=\grphi{}+{\bf J}/P_s$ and $\left.\dot {\bf x}\right|_{downhill}=-\grphi{}+{\bf J}/P_s$.
These paths correspond to gradient ascent and descent, similarly to  equilibrium. However, contrary to this case, the system also flows along the direction given by the probability current and therefore visit different portions of the phase space depending whether it is moving uphill or downhill \footnote{A simple and instructive case, suggested to us by J.-P. Bouchaud, is provided by a force field ${\mathbf F}=-{\boldsymbol \nabla} V+{\boldsymbol \nabla}\times {\mathbf A}$ such that rotational and irrotational contributions are orthogonal for any value of $\mathbf x$. This property means that the rotational part makes the system
flows along the iso-potential surfaces. In this case one can obtain explicitly several results discussed in the text. First, it is easy to check that the steady state distribution is given by the Boltzmann-like distribution $e^{-V/T}$ despite the fact that the system is out of equilibrium. Second, the steady state
probability current reads ${\bf J}/P_s={\boldsymbol \nabla}\times {\mathbf A}$ . In this case the decomposition 
of downhill and uphill trajectories discussed previously is very natural and follows directly from the definition of $\mathbf F$.}.  \\
Another interesting finding that differentiates activation in and out of equilibrium 
concerns the work done on the system by the force field: $W=\int dt\, \dot {\bf x}\cdot {\bf F}$.
We can compute $W$ along the two trajectories and obtain that, in both cases, it can be split into two contributions:
$W=T\int dt\,{\bf J}^2/(TP_s^2)-(\phi(x_f)-\phi(x_i)) $. 
If we compute the work done in a closed cycle, we obtain that only the first term, related to the current, is always dissipated while the second term vanishes. The system indeed releases to the bath an amount $\Delta \phi$ when it gradient descends and it absorbs the same quantity during the gradient ascent. 
Using the result in \cite{seifert}, which states that the total entropy produced along a given trajectory reads $\Delta S=\int dt\,{\bf J}^2/(TP_s^2)$,
one can recast the previous equation for $W$ in a form resembling the first law of thermodynamics: $W=T\Delta S-\Delta \phi$.
The main differences with respect to the first law are that the previous expression is valid for a given trajectory and that the internal energy, which does not exist for a driven system, is replaced by the out of equilibrium potential $\phi$.\\        
\begin{figure}
\includegraphics[width=.95\columnwidth]{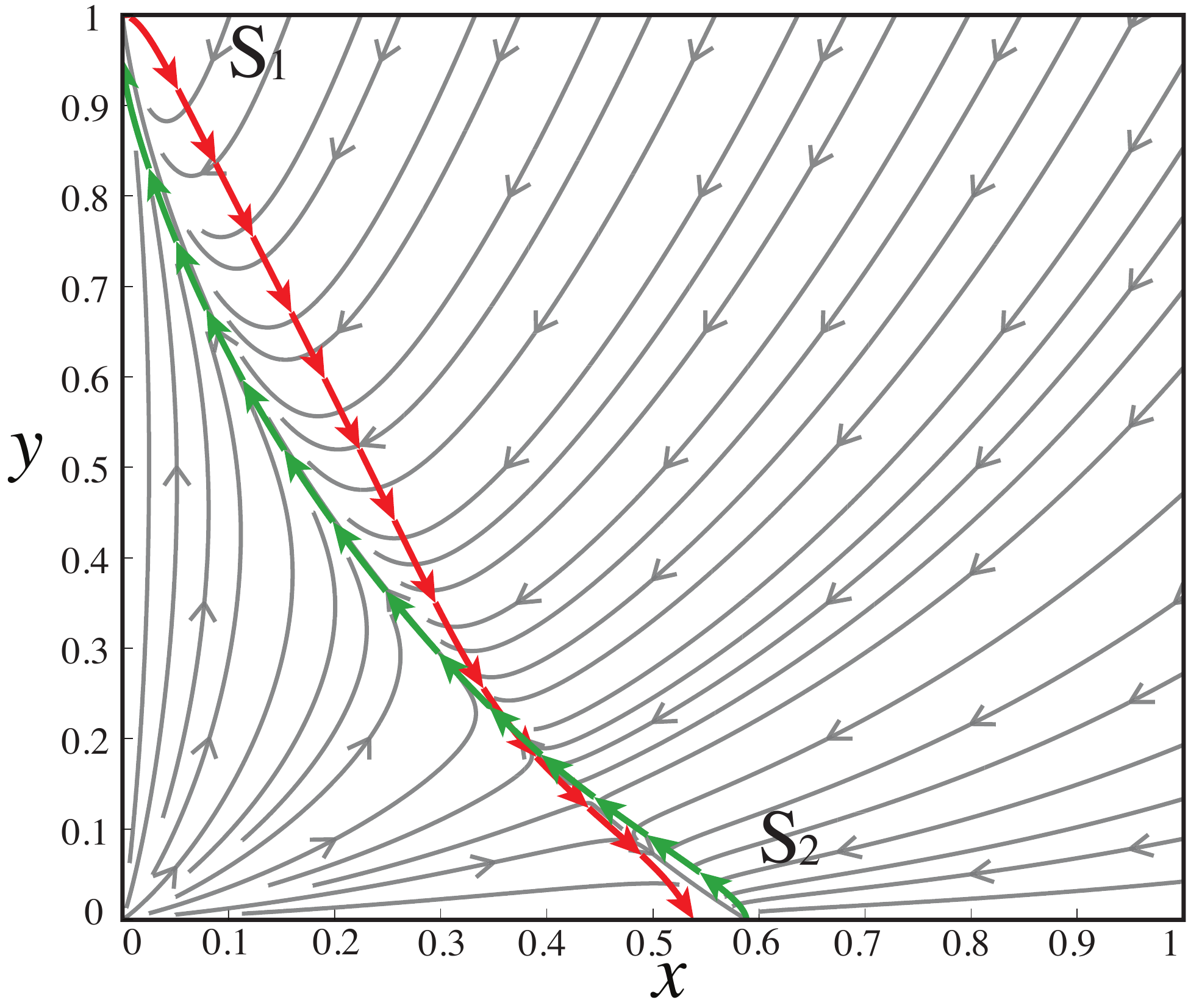}
\caption{Saddle point trajectories in the field of forces ${\bf F}(x,y)$ with the choice of the parameters $\epsilon=1.7$, $f=1.12$,  $\psi=1.1$ and initial condition $x_i=S_1, |\hat{x}_i|=0.0032$ (red) and $x_i=S_2, |\hat{x}_i|=0.0032$ (green). The plot shows clearly that uphill and downhill paths visit distinct regions of the configuration space. }
\label{saddlepoints}
\end{figure}
Let us finally discuss the most general physical case we are 
able to treat, which is characterized by the following system of stochastic equations: 
\begin{eqnarray}\label{general}
m\ddot{x}_\alpha+&&\sum_\beta \int^t_{-\infty}dt'\gamma_{\alpha,\beta}(t-t')\dot{x}_{\beta}=F_\alpha+\,\,\eta_\alpha\nonumber\\
&&\langle\eta_\alpha(t)\eta_\beta(t')\rangle=2T \tilde \gamma_{\alpha,\beta}(t-t')H({\bf x})
\end{eqnarray}
where $\gamma_{\alpha,\beta}(t-t')$ is a function related to retarded friction, symmetric with respect to the interchange of $\alpha,\beta$ and equal to zero for $t<t'$ and $\tilde \gamma_{\alpha,\beta}(t-t')$ is a
symmetric positive-definite operator  associated to non-Markovian noise \footnote{The other property verified by 
$\gamma_{\alpha,\beta}(t-t')$ is that $\gamma_{\alpha,\beta}(t-t')+\gamma_{\alpha,\beta}(t'-t)$ is a symmetric positive-definite operator, see the SI text.}. Physically, eq. (\ref{general}) corresponds to a system which 
undergoes Netwonian dynamics in a non-potential force field and is coupled to an out of equilibrium thermal bath (only in the case $H=1$ and for $\gamma_{\alpha,\beta}(t-t')+\gamma_{\alpha,\beta}(t'-t)=2\tilde \gamma_{\alpha,\beta}$ the bath is at equilibrium). 
The main trick we used to analyze this case consists in reducing the problem to the one we already solved. This is done by showing that 
the general stochastic equation above can be rewritten in terms of over-damped Langevin equations characterized by white noise
in an extended configuration space where new extra variables are introduced: the inertial term is handled by introducing the momentum, ${\bf p}=m\dot {\bf x}$; whereas retarded friction and non-Markovian noise can be represented 
as the result of integrating out a bath of harmonic oscillators linearly coupled to the system and evolving
by an over-damped Langevin equation characterized by white noise \cite{zwanzig}. More details and the explicit calculations can be found in the SI text \footnote{See SI text}.

In conclusion, we provided a general and comprehensive analysis of the problem of activation out of equilibrium. 
We showed that the Arrhenius law holds for a very large class of out of equilibrium systems provided that the energy function is replaced by the non-equilibrium potential $\phi$. The most important difference with the equilibrium case is that the noise-activated paths are no longer related by a simple time reversal transformation: uphill and downhill paths visit different regions of the configuration space.
We characterized these trajectories by obtaining their explicit expression, by identifying their irreversible nature and by unveiling how they are related to a generalization of the time-reversal transformation. We illustrated our results in an explicit example borrowed from biology where 
{\em equilibrium} is not even a well defined concept. We envision many possible interesting applications
in different fields ranging from physics to economics, where equilibrium is a limiting assumption.   

\acknowledgments
We thank J.-P. Bouchaud, J. Kurchan, M. Marsili and A. Silva for helpful discussions. 
GB acknowledges support from ERC grant NPRG-GLASS. SB acknowledges CEA-Saclay for hospitality.

%

\end{document}